\documentclass[conference]{IEEEtran}
\IEEEoverridecommandlockouts
\usepackage{cite}
\usepackage{amsmath,amssymb,amsfonts}
\usepackage{algorithmic}
\usepackage{graphicx}
\graphicspath{ {Figures/} }
\usepackage{textcomp}
\usepackage{xcolor}
\usepackage{hyperref}
\usepackage{algorithm}
\usepackage{algorithmic}
\usepackage{booktabs}
\usepackage{multirow}
\usepackage[separate-uncertainty,tight-spacing=true]{siunitx}
\usepackage{blindtext}

\def\BibTeX{{\rm B\kern-.05em{\sc i\kern-.025em b}\kern-.08em
    T\kern-.1667em\lower.7ex\hbox{E}\kern-.125emX}}
\begin{document}

\title{Trust Management in Decentralized IoT Access Control System
}

\author{
\IEEEauthorblockN{
    Guntur Dharma Putra\IEEEauthorrefmark{1},
    Volkan Dedeoglu\IEEEauthorrefmark{2},
    Salil S. Kanhere\IEEEauthorrefmark{1},
    and Raja Jurdak\IEEEauthorrefmark{3}
}
\IEEEauthorblockA{
    \IEEEauthorrefmark{1}UNSW, Sydney
    \IEEEauthorrefmark{2}CSIRO Data61, Brisbane
    \IEEEauthorrefmark{3}QUT, Brisbane \\
    \{gdputra, salil.kanhere\}@unsw.edu.au, volkan.dedeoglu@data61.csiro.au, r.jurdak@qut.edu.au
}
}

\maketitle

\begin{abstract}
Heterogeneous and dynamic IoT environments require a lightweight, scalable, and trustworthy access control system for protection from unauthorized access and for automated detection of compromised nodes.
Recent proposals in IoT access control systems have incorporated blockchain to overcome inherent issues in conventional access control schemes. However, the dynamic interaction of IoT networks remains uncaptured.
Here, we develop a blockchain based Trust and Reputation System (TRS) for IoT access control, which progressively evaluates and calculates the trust and reputation score of each participating node to achieve a self-adaptive and trustworthy access control system. Trust and reputation are explicitly incorporated in the attribute-based access control policy, so that different nodes can be assigned to different access right levels, resulting in dynamic access control policies.
We implement our proposed architecture in a private Ethereum blockchain comprised of a Docker container network. We benchmark our solution using various performance metrics to highlight its applicability for IoT contexts.
\end{abstract}



\section{Introduction}
\label{sec:introduction}
Access control, or authorization, is designed to limit the actions and operations an authenticated user can perform on a computer system~\cite{sandhu1994access}. By enforcing restrictions on user actions, 
access control aims to prevent security breaches that may lead to the exposure of important resources to unwanted adversaries. An example of a popular and widely used authorization scheme is OAuth~\cite{hardt2012oauth}, which provides secure delegated third-party access. In this scheme, access to resources is protected by a centralized entity on behalf of the resource owner.

Within the context of IoT, conventional access control must be tailored to match the specific requirements of IoT. It is commonly known that IoT devices are resource-constrained, e.g. limited energy, memory capacity, and computational power. Hence, access control mechanisms designed for IoT should avoid using computationally expensive and memory intensive operations. Instead, access protocols based on simpler signature and hashing operations are required.
Furthermore, a typical IoT deployment involves a large number of sensor and actuator nodes, making scalability another key issue to consider. IoT access control protocols must also be flexible to accommodate the dynamic nature of the system and heterogeneity of the devices.

There has been an increasing trend towards utilizing blockchain in access control for IoT~\cite{Novo2018, Pinno2018, Jiang2018, Ding2019, Andersen2019, DiPietro:2018:BTS:3205977.3205993, DORRI2019180,8894097}. Researchers argue that salient features of blockchain, for instance decentralization and immutability, have the potential to overcome unsolved inherent issues in conventional access control, such as single point of failure and lack of transparency. For instance, in~\cite{DORRI2019180}, blockchain is used as a distributed and immutable storage of access control policies. Smart contracts in blockchain are also used to deliver secure and transparent access control using static access right validations defined in access policies~\cite{Jiang2018}.
However, little attention has been aimed towards incorporating trust management in decentralized access control. In fact, due to scalability and dynamic interactions of IoT nodes in a network with intensive data streams and interactions, it is imperative to build a dynamic and flexible authorization system. Ideally, an automated trust management or Trust and Reputation System (TRS) would observe the behavior of every participating node by progressively quantifying its behavior to a trust or reputation score. As a result, benign and misbehaving nodes are associated with high and low reputation score, respectively. The TRS would extend current access control scheme, where trust and reputation are incorporated in the access control policy to achieve dynamic authorization system.
To optimally utilize blockchain technology, the TRS should leverage the advantages offered by blockchain, such as pseudonymity, transparency, secure collaborative computation, and non-repudiation in transactions.

In this paper, we incorporate trust management in blockchain-based IoT authorization systems to achieve decentralized, secure, and trustworthy access control. We use smart contracts to replace centralization and to deliver transparent processing. In the context of large-scale IoT access control, the requester and requestee are best represented and identified as attributes rather than actual identities~\cite{8752021}. We model our access control scheme as an Attribute-Based Access Control (ABAC)~\cite{sandhu1994access}. Trust and reputation values are the additional attributes for access request validation. By combining trust and reputation with inherent attributes of IoT nodes, e.g., hardware and software/firmware specification, and device ownership, we can prevent colluding nodes from building false reputations to gain access to protected resources. In our definition, trust is a subjective view of a node's behavior based on its previous interactions, independently calculated by each node in the network. On the other hand, reputation is a global view of a node's past behavior from aggregated trust relationships in other nodes~\cite{DiPietro:2018:BTS:3205977.3205993}, calculated transparently by the smart contracts. We adopt Gompertz function for trust value calculation~\cite{Huang2010} and formulate a reputation score for global trust computation. Progressive trust and reputation evaluation may effectively help to detect and eliminate malicious or compromised nodes in the network~\cite{MohamadNoor2019}. By incorporating blockchain technology and signature-based detection, our proposed TRS is designed to be robust against manipulations of unauthorized or malicious entities. Furthermore, we also incorporate blockchain events as a way to propagate and notify peers in case of access violations. The TRS is designed to be blockchain-agnostic and can be implemented in any blockchain instantiation with adequate support for smart contract execution.

In summary, the contributions of this paper are:
\begin{itemize}
    \item We design a lightweight TRS for decentralized IoT access control. We use smart contracts to ensure the transparency and reliability of the system. By progressively evaluating the trust and reputation of nodes, TRS captures the dynamics of an IoT deployment, and detects and eliminates corrupt or malicious nodes in the network.
    \item We incorporate blockchain events as a method to automatically notify peers in case of access control violations. As the events propagate, proper actions and mitigation can be executed to prevent unwanted consequences from corrupt or malicious nodes.
    \item We develop a proof-of-concept implementation of our proposed framework in a private Ethereum network. To demonstrate the applicability of our framework in the IoT context, we use four benchmark metrics, e.g., latency, gas consumption, and reputation and trust evolution.
\end{itemize}

The rest of the paper is organized as follows. In Section~\ref{sec:related-works}, we summarize related works. Section~\ref{sec:architecture} describes our proposed architecture. We present our proof-of-concept implementation in Section~\ref{sec:performance-evaluation}. Section~\ref{sec:discussion} discusses the limitations of our framework and Section~\ref{sec:conslusion-future-work} concludes the paper.


\section{Related Works}
\label{sec:related-works}
To date, several works have proposed decentralization in IoT access control, which primarily aim to remove trusted and centralized processing. Pinno et. al.~\cite{Pinno2018} proposed a fully decentralized access control in IoT with four separate blockchains that work together, namely context, accountability, relationships, and rules. Information about entities and authorization rules is stored as blockchain transactions, in which three types of access control mechanisms are supported, i.e., Access Control List (ACL), capability-based, and attribute-based access control. In~\cite{Ding2019}, authors proposed decentralized attribute-based authorization scheme, in which authorization and revocation of attributes are recorded in blockchain transactions. Access control is enforced by the resource owner and attributes validation is performed by using blockchain search.
However, both works~\cite{Pinno2018, Ding2019} do not optimally utilize all capabilities offered by blockchain and primarily operate on off-chain processes, e.g., no smart contracts are implemented. The ACL also does not perform optimally on a network with dynamic and large number of connected devices, as the ACL must be manually updated to keep track of joining and leaving nodes.

Recently, a number of studies have incorporated smart contracts to deliver decentralization of access control in IoT. In~\cite{Jiang2018}, the authors replaced central authenticator with smart contracts, including access control contracts, a single judge contract, and a single register contract. An ACL is stored in the smart contract to record subject-object-action relationships. The smart contract authenticates and authorizes a requester by checking whether the requester has the access right according to the ACL. Novo proposed to utilize a single smart contract to enforce access control in a private blockchain~\cite{Novo2018}, in which all permitted access operations are recorded in the smart contract. The managers of IoT devices are responsible to authorize requesters by calling a function in the smart contract to validate the requests against predefined ACL. In~\cite{Andersen2017}, authors introduced the concept of Delegation of Trust (DoT) as a proof of authorization to certain resources in a namespace stored in a blockchain. DoTs are issued by the smart contracts, which incorporate identity based encryption.
All above works~\cite{Jiang2018, Novo2018, Andersen2017} use smart contracts to replace centralized processing and enable transparent access control, but all are unable to capture the dynamics that are inherent in large IoT ecosystems.

Recent research has explored the inclusion of trust in IoT access control. Bernabe et. al.~\cite{BernalBernabe2016} built a trust model as an additional layer of security in a Trust-aware Access
Control Mechanism for IoT (TACIoT) system. The trust is calculated in a fuzzy fashion by a central trust manager and is used for validation during each authorization process. Mahalle et. al.~\cite{Mahalle2013} proposed a Fuzzy Trust Based Access Control (FTBAC) that maps trust scores to certain access permissions of a resource with the principle of least privilege. Each node in the network periodically calculates trust score of its neighboring nodes based on experience, knowledge, and recommendation. In~\cite{DiPietro:2018:BTS:3205977.3205993}, authors proposed a three-way handshaking protocol for access control, in which the requester and requestee agree upon certain terms and obligations on accessing a resource. The terms and obligations examine the requestee global reputation, which is stored on the blockchain.
Although trust is incorporated in~\cite{BernalBernabe2016, Mahalle2013} as an additional layer of security, trusted and centralized processing units are still present.
The blockchain based solution~\cite{DiPietro:2018:BTS:3205977.3205993} is rather complex and involves complicated and inefficient mechanism, which renders it to be unsuited for IoT.

In summary, previous studies on decentralized IoT access control/authorization are unable to address dynamic behavior of IoT deployment. Furthermore, little attention has been devoted to provide a secure trust management for access control in IoT.
In this paper, we propose a TRS in decentralized IoT access control system to provide trustworthy access control and automated malicious node detection. 


\section{Proposed System Architecture}
\label{sec:architecture}
In this section, we outline our proposed system architecture and describe the access control mechanism along with the blockchain-based TRS.
We propose blockchain network as the main actor for authorization and trust management to achieve a secure, trustworthy, and transparent access control.

\begin{figure}[!t]
\centering
\includegraphics[width=0.47\textwidth]{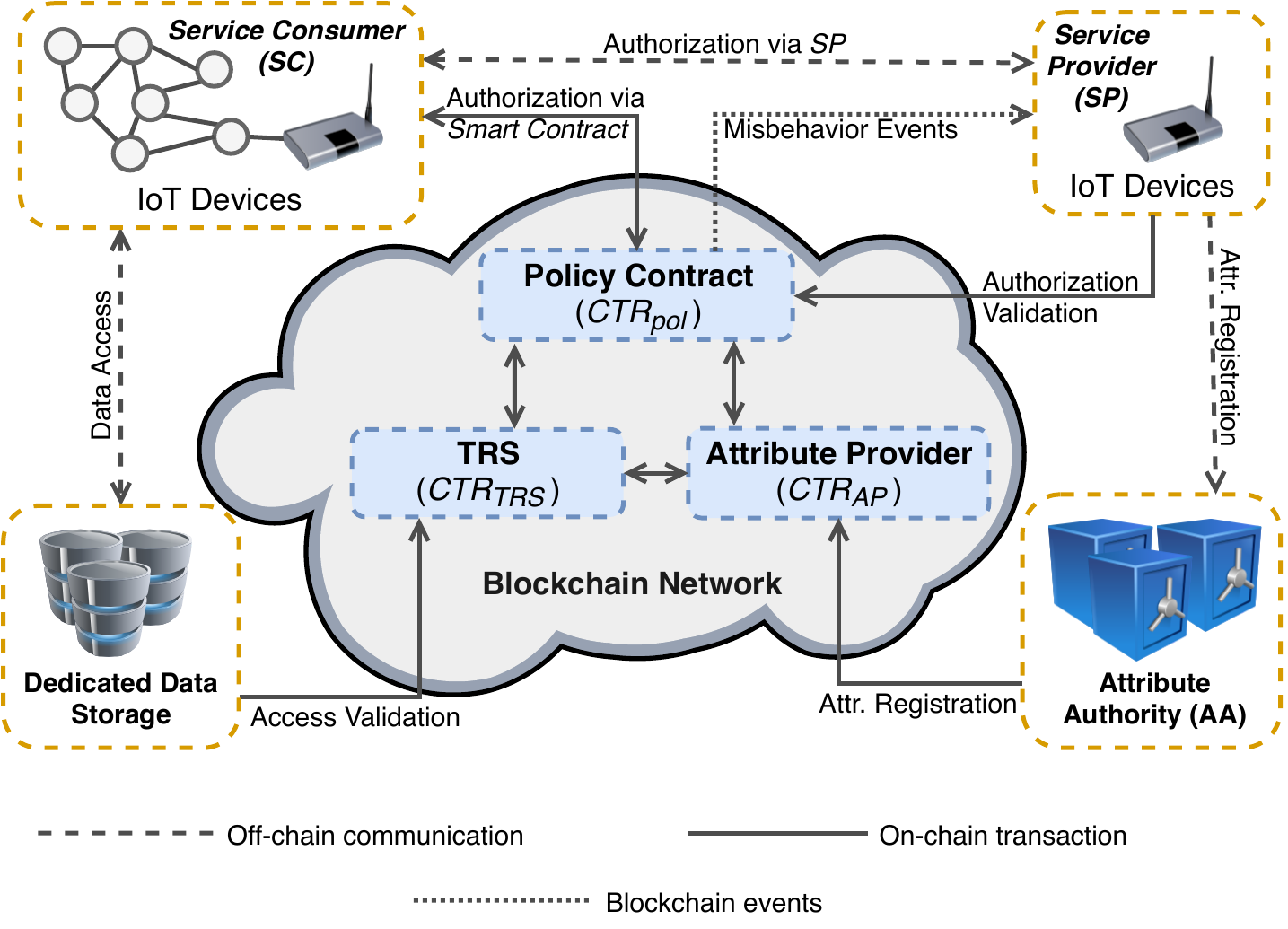}
\caption{Architecture overview.}
\label{fig:architecture-overview}
\end{figure}
    
    \subsection{Architectural Components}
    \label{sub:architectural-components}
    Fig.~\ref{fig:architecture-overview} depicts the main components of our architecture which include: IoT devices, data storage, attribute authority, and the blockchain network.
    We use a smart building application as a use case, where we expect interactions from known, i.e., devices controlled by building manager, and unknown devices, i.e., ones from third-party vendors.
    However, we design our architecture to be application agnostic and capable of being adopted to other IoT scenarios. Next, we describe these components in detail.
    
    \subsubsection{IoT devices}
    \label{subsub:iot-dev}
    The IoT devices correspond to any nodes in an IoT deployment, such as smart locks, security cameras, smart meters, and other connected devices deployed in a smart building.
    The nodes include the Service Providers ($SP$s), e.g., temperature sensors, which own resource $r$ controlled as per access policy $P$, and the Service Consumers ($SC$s), e.g., devices from third party vendors, which consume resource $r$. IoT nodes with limited computational capability, connect to the blockchain network via gateways, while nodes with adequate computational power, i.e., to perform asymmetric cryptography, connect to the blockchain network directly~\cite{elktechcrunch}. Both $SP$ and $SC$ are identifiable by their Public Keys ($PK$s).
    
    \subsubsection{Data storage}
    \label{subsub:data-storage}
    An $SP$ may store data temporarily in its local memory, while a dedicated off-chain data storage stores high volume of data over a longer timespan. To access the data, $SC$ sends an access request to the dedicated data storage after obtaining an authorization token. We describe the access control scheme in more detail in Section~\ref{sub:access-control-scheme}.
    
    \subsubsection{Attribute Authority}
    \label{subsub:attribute-authority}
    In our architecture, an Attribute Authority ($AA$) is in charge of issuing legitimate attributes to each IoT node. In a large scale IoT deployment, IoT nodes are best described in terms of attributes rather than their actual identities~\cite{8752021}. Thus, we store IoT devices' attributes (e.g., sensor types, role, and hardware specifications) on the blockchain rather than their real identities. Note that, in our architecture, the $AA$ is not responsible for policy assignment and access control mechanism. The policy assignment is managed by $SP$, while access control mechanism is enforced by the smart contract. Within the context of a smart building, the $AA$ is managed by the building manager who manages the building specification in which technical sensor deployments are described. The $AA$ is identified by its Public Key $PK_{AA}$. We discuss attributes in more detail in Section~\ref{sub:attributes-policies}.
    
    \subsubsection{Blockchain network}
    \label{subsub:blockchain-network}
    We use a blockchain network for decentralizing access control, by managing the associated transactions and running the access control logic. This provides higher scalability, better security, and eliminates a single point of trust and failure compared to common centralized access control mechanisms.  Furthermore, we compose three main smart contracts, namely Attribute Provider ($CTR_{AP}$), Trust and Reputation System ($CTR_{TRS}$), and Policy Contract ($CTR_{pol}$), to automate the execution of attribute validation, trust calculation, and access policy validation, respectively. We design our architecture to be blockchain agnostic, which complies with any blockchain implementation that supports smart contract execution.

    \subsection{Attributes and Policies}
    \label{sub:attributes-policies}
    An $SC$ must register its attributes to the $AA$ prior to requesting access to a resource. An attribute ${ \alpha }_{ i }^{ m }$ of $SC_i$ is denoted as:

    \begin{equation}
        { \alpha }_{ i }^{ m }=\left< key, type, val \right> 
    \end{equation}
    
    \noindent where $key$, $type$, and $val$ correspond to the name, type, and value of the attribute, respectively. An $SC_i$ may have a set of $M$ attributes, $A_i=\{\alpha_i^1, \alpha_i^2, ..., \alpha_i^M\}$, which defines its characteristics. $SC$ uses an off-chain secure channel to send attribute registration request to $AA$, and sign the request with its secret key $SK_{SC}$ for authentication purpose. After receiving the request, $AA$ verifies the attributes according to the smart building specifications to prevent attribute forgery. Once the verification is completed, $AA$ stores the attributes in the blockchain by initiating a registration transaction:

    \begin{equation}
        { TX }_{ reg }=\left[ { A }_{ i }|{ Sig }_{ SC }| timestamp | {  Sig }_{ AA } \right] 
    \end{equation}
    
    \noindent where ${ Sig }_{ SC }$ and ${ Sig }_{ AA }$ are the signatures of $SC$ and $AA$, respectively. ${ Sig }_{ AA }$ is proof that the transaction was initiated by the $AA$. The ${ TX }_{ reg }$ transaction is handled by $CTR_{AP}$ and is stored on the blockchain as a key value pair. Note that, we assume that there is no privacy issue associated with these attributes, as they are assumed to be publicly available in the smart building specifications and the $AA$ does not store a direct binding between entities and their real identities.
    
    
    
    
    
    
    
    To explicitly define the eligibility for accessing a resource, a Service Provider ($SP$) needs to define an access policy $P$, which expresses a Boolean rule set that can evaluate several attributes. The $SP$ has the authority to define, update, and revoke the access policies. An $SP$ may allow an $SC$ to perform a set of actions $\tau\subset\{read,write,stream\}$ on a resource $r$ in a context $c$, if $SC$ satisfies policy ${ P }_{ r,c }$, which is expressed as:
    
    \begin{equation}
        { P }_{ r,c }=\left< A_p , c_p, \tau_p, Rep_{min} \right> 
    \end{equation}
    
    \noindent where $A_p=\{ { \alpha  }_{ p,1 },..,{ \alpha  }_{ p,n }\}$ is the set of required attributes, $c_p=\left<t, l\right>$ is the set of contexts (time and access throughput limit) and $Rep_{min}$ is the minimum reputation value. The $SP$ is required to initiate transaction ${ TX }_{ pol }$ to create and store the policy in  $CTR_{pol}$, which is defined as:
    
    \begin{equation}
        { TX }_{ pol }=\left[ r_p|\tau_p |A_p|Rep_{min}|c_p|timestamp|Sig_{ SP } \right] 
    \end{equation}
    
    \noindent where $timestamp$, and $Sig_{ SP }$ correspond to the policy creation time, and signature of $SP$, respectively.

    \subsection{Trust and Reputation System}
    \label{sub:trust-reputation-management}
    Generally, a Trust and Reputation System (TRS) evaluates the trustworthiness of participating nodes in a particular system to capture the dynamics of the network. Our TRS also aims to enhance access control security by detecting and eliminating malicious or compromised nodes.
    We refer to trust as a subjective local belief that a node will behave normally, while reputation refers to a global view of a node's previous behavior. In this section, we describe how our lightweight and transparent TRS mechanism works.
    
    The trust value of an $SC_i$ by $SP_j$, denoted as $T_{i,j}$, is calculated by $SP_j$ from its previous interactions with $SC_i$, which may be positive or negative. Positive interaction corresponds to honest action of $SC_i$, while negative interaction is otherwise. If $SP_j$ has no previous interaction with $SC_i$, $T_{i,j}$ will be assigned to a default value of $0$. The trust value is calculated when $SC_i$ requests an access to a resource $r$ of $SP_j$. Note that, $SP_j$ is responsible for storing and progressively evaluating trust values across different nodes in the network.
    
    We adopt the trust concept in~\cite{Huang2010} to calculate the trust value of $T_{i,j}$, in which the Gompertz function is used to model the trust growth. As in real-life social interactions, trust increases gradually by positive experience in interactions and drops significantly by a negative experience. We also consider recent interactions to be more relevant than previous interactions. In addition to Gompertz growth function, we introduce an ageing function, which assigns higher weights for recent interactions and lower weights for past interactions. The ageing function of $n$-th interaction ${ I }_{ n }$ is formulated as follows: 
    
    \begin{equation}
    \label{eq:ageing-function}
        { I }_{ n }=\sum _{ i=1 }^{ n }{ { \gamma  }^{ \left( n-i \right)  }{\delta}_{i}  } 
    \end{equation}
    
    \begin{equation}
        {\delta}_{i} =\begin{cases} {\delta}_{pos} \quad if \, positive \, interaction \\ {\delta}_{neg} \quad otherwise \end{cases}
    \end{equation}
    
    \noindent where ${ \gamma  }^{ \left( n-i \right)  }$ is the ageing parameter, where $0 < \gamma < 1$, and  $\delta_i$ corresponds to the weight of interaction with ${\delta}_{neg} < 0$, and ${\delta}_{pos} > 0$. Note that, we set ${\delta}_{pos} < \left| {\delta}_{neg} \right| $ so it is harder for the $SC$ to build trust than to lose it.
    
    The output of Eq.~\ref{eq:ageing-function} is fed to the Gompertz growth function that limits the trust value from $0$ (not trusted) to $1$ (fully trusted). The formula to calculate trust value ${T}_{i,j}$ is formulated as follows:
    
    \begin{equation}
    \label{eq:gompertz-function}
        {T}_{i,j}\left( { I }_{ n } \right) =a.{ e }^{ -b.{ e }^{ -c.{ I }_{ n } } }
    \end{equation}

    \noindent where $a$, $b$, and $c$ are the asymptote, the displacement parameter along \texttt{x}-axis, and the growth rate, respectively.
    
    In contrast with trust, which is a subjective view of nodes' behavior, reputation is a global view, which is built by aggregating $SC_i$ interactions from multiple $SP$s. The reputation of an $SC_i$, $Rep_i$, is recorded in the blockchain and is progressively evaluated by the trust and reputation smart contract $CTR_{TRS}$, when trustworthy interaction is recorded or malicious behavior is reported.
    
    In our scheme, reputation value must also consider the variation of communicating counterparts of $SC_i$. That way, node $SC_i$ may not be able to conceivably build deceptive reputation by only interacting with a single $SP$. To build high reputation value, $SC_i$ must have a trustworthy transaction history from multiple $SP$s. Similar to trust calculation, we consider recent interactions to be more relevant than previous ones. The reputation of $SC_i$ on its $n$-th interaction is calculated as follows:
    
    \begin{equation}
    \label{eq:reputation-formula}
        { Rep }_{ i,n }=\left( \sum _{ t=1 }^{ n }{ \beta_t.{ \lambda  }^{ \left( n-t \right)  } }  \right) \ln { \left( { N }_{ peers } \right)  }  
    \end{equation}
    
    \begin{equation}
        \beta_t =\begin{cases} {\beta}_{pos} \quad if \, positive \, interaction \\ {\beta}_{neg} \quad otherwise \end{cases}
    \end{equation}
    
    \noindent where $\beta_t$ is the weighting parameter with $\beta_{neg} < 0$ and $\beta_{pos} > 0$, ${ \lambda }^{ -\left( n-t  \right)  }$ is the decaying parameter (i.e., $0 < \lambda < 1$), and ${ N }_{ peers }$ represents number of interacted peers.
    We set ${\beta}_{pos} < |{\beta}_{neg}|$ so that it is more difficult to gain reputation than to lose it.
    When a new $SC_k$ joins the network, $CTR_{TRS}$ assigns a default reputation value of $0$, i.e., $Rep_{k,0}=0$, which implies that $SC_k$ can only access resources with $Rep_{min}=0$.

    
    \subsection{Access Control Scheme}
    \label{sub:access-control-scheme}
    Access control policy is enforced by $CTR_{pol}$ in the blockchain. $CTR_{pol}$ evaluates an access request on whether the set of requester's attributes satisfy certain Boolean attribute expressions in the policy for resource $r$. Note that, trust and reputation values are also part of the attributes. The process starts when an $SC_i$ requests access for a resource $r_{j,k}$ of $SP_j$.
    Depending on the access policy $P_{r,c}$, $SC_i$ is required to send an access request directly to the $SP_j$ (Section~\ref{subsub:access-via-sp}) or via the smart contract $CTR_{pol}$ (Section~\ref{subsub:access-via-smart-contract}) for a relaxed setting.
    Upon successful validation, $CTR_{pol}$ issues an access token containing access details (e.g., token expiration and access limits). The token can be used to re-access the resource prior to expiration time, without the need of repeating the access request process, which improves the latency. Any violations of access request are also recorded in blockchain by $CTR_{TRS}$. The following subsections elaborate on the system setup, access control mechanism, and malicious behavior reporting in more detail.
    
    
    \subsubsection{System Setup}
    \label{subsub:system-setup}
    The smart building manager initially deploys all mandatory smart contracts, e.g., Attribute Provider ($CTR_{AP}$), Trust and Reputation Contract ($CTR_{TRS}$), and Policy Contract ($CTR_{Pol}$) to the blockchain. Recall that an $SP$ may refer to an IoT gateway $SP_{gt}$ or an independent IoT node $SP_{nd}$. Initially, each $SP_{gt}$ performs device and resource discovery to map associated IoT nodes and their specific resources. The process is carried out according to specifications in low level routing and networking protocols for IoT~\cite{Islam2019}. $SP_{gt}$ stores the mapping results $G(r, n_{id})$, where $r$ is the resource and $n_{id}$ is the node ID, in the local resource database. Both $SC_{gt}$ and $SC_{nd}$ may need to coordinate with data storage for long term storage of the data. Once available resources are identified, both $SP_{gt}$ and $SP_{nd}$ construct and register corresponding policies, described in Section~\ref{sub:attributes-policies}. By the end of the system setup process, all smart contracts are deployed, and policies are constructed and stored in the blockchain.

    \subsubsection{Authorization via service provider}
    \label{subsub:access-via-sp}
\begin{figure}[!t]
    \centering
    \includegraphics[width=0.47\textwidth]{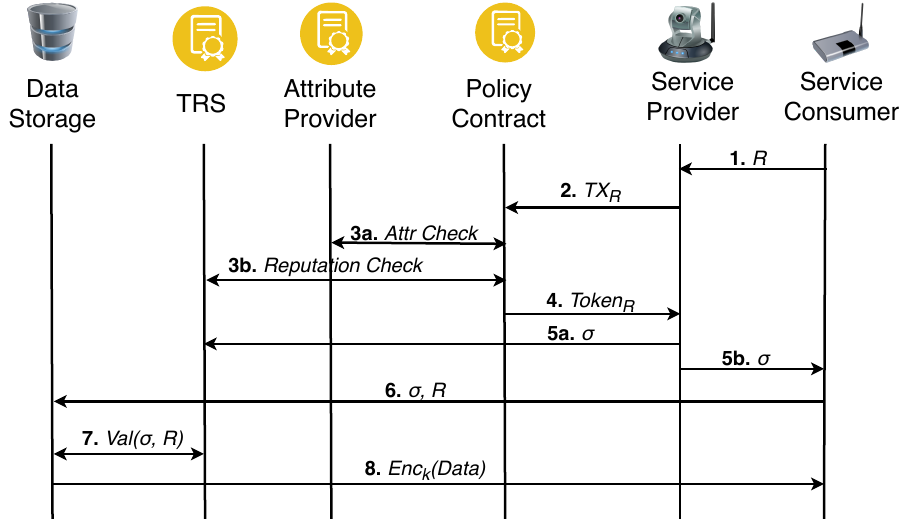}.
    \caption{Resource access via Service Provider.}
    \label{fig:access-via-sp}
    \end{figure}

    We show the process of access request via a service provider in Fig.~\ref{fig:access-via-sp}. The process involves getting the access token from the $SP$, and collecting the data from the data storage. In this scenario, we assume that the $SC_i$ has already obtained information about $SP_j$ and the resource identifier. We also assume that both $SP$ and $SC$ have sufficient computational power to perform asymmetric cryptography.
    
    \begin{itemize}
        \item Step 1: The $SC_i$ sends a request to $SP_j$ for an access to a resource $r$ with action $\tau$ by sending a message $R$ in a secure channel, which is defined as:
        
        \begin{equation}
            R=\left< r, \tau, S, { Sig }_{ SC } \right> 
        \end{equation}
        
        where $r$ is the resource identifier, $S={Enc}_{{PK}_{SP}}(k)$ is a session key $k$ encrypted by public key of service provider ${PK}_{SP}$, and ${ Sig }_{ SC }={Sig}_{{SK}_{Sc}}(Hash(R))$ is the signature of the request message, which is used for authentication. 
        
        \item Step 2: The $SP$ forwards $R$ to $CTR_{pol}$ for policy validation by initiating a transaction $TX_R$, which is defined as:
    
        \begin{equation}
        \label{eq:tx-r}
            { TX }_{ R }=\left[ R | Sig_{ SP } \right] 
        \end{equation}
        
        where $Sig_{ SP }$ is the signature of the $SP$.
        
        \item Step 3a and 3b: To validate the request, $CTR_{pol}$ obtains the attributes and reputation of $SC_i$ by calling attribute check to $CTR_{AP}$ (3a) and reputation check to $CTR_{TRS}$ (3b). Access request validation is performed according to Algorithm~\ref{alg:access-control-mechanism}.
        
        \item Step 4: Upon successful validation, $CTR_{pol}$ issues $Token_{R}$, which is defined as:
        
        \begin{equation}
            Token_{R}=\left<{ Rep }_{ i }, {Exp}_{R}, l, t\right>
        \end{equation}
        
        where ${ Rep }_{ i }$ is the reputation of $SC_i$, ${Exp}_{R}$ is the token expiration time, $l$ is the access throughput limit, and $t$ is timestamp. $CTR_{pol}$ sends $Token_{R}$ to $SP_j$ for final access decision. If $CTR_{pol}$ observes repeated access, in which attributes or reputation value does not satisfy policy $P_{r}$, $CTR_{pol}$ reports this as a malicious activity to the $CTR_{TRS}$.
        
        \item Step 5a and 5b: The $SP_j$ finalizes the token based on a decision function $g$:
        
        \begin{equation}
        \label{eq:access-final-decision}
            { Decision }_{ i,k }=g[{ T }_{ i,j },{ Rep }_{ i }]
        \end{equation}
        
    Note that, in this scheme, $SP_j$ has the full authority to grant or deny the access request. When function $g$ grants access to $SC_i$, $SP_j$ finalizes the process by signing $Token_R$ using its private key, $\sigma=Sig_{{SK}_{SP_j}}({Token}_{R})$ and storing it in the blockchain via $CTR_{TRS}$ and sending $\sigma$ to $SC_i$. At this point, $SC_i$ is authorized to access the resource.
        
        \item Step 6, 7, and 8: To collect the data from resource $r$, $SC_i$ sends a message $\left<\sigma, R\right>$ to the data storage. Data storage validates $\left<\sigma, R\right>$ to $CTR_{TRS}$ to prevent access token forgery. Upon successful validation, data storage sends ${Enc}_{k}(Data)$, data encrypted by session key $k$ to $SC_i$.
    \end{itemize}
    
    \begin{algorithm}
    \caption{Access Control Check}
    \label{alg:access-control-mechanism}
    \begin{algorithmic}[1]
    \renewcommand{\algorithmicrequire}{\textbf{Input:}}
    \renewcommand{\algorithmicensure}{\textbf{Output:}}
    \REQUIRE $P_{r,c}$, $TX_R$, $Rep_i$, and $A_i$
    \ENSURE  $Token_R$ or $\emptyset$
    \STATE $authorized \leftarrow 0 $
    \IF{$A_p \subset A_i$}
        \IF{$\tau \subset \tau_p$}
            \IF{$Rep_i \ge Rep_{min}$}
                \STATE $authorized \leftarrow 1 $
            \ENDIF
        \ENDIF
    \ENDIF
    \IF{$authorized$}
        \RETURN $Token_R \leftarrow \left<{ Rep }_{ i }, {Exp}_{R}, l, t\right>$
    \ELSE
        \RETURN $\emptyset$
    \ENDIF
    \end{algorithmic}
    \end{algorithm}

    \subsubsection{Authorization via Smart Contract}
    \label{subsub:access-via-smart-contract}
\begin{figure}[!t]
    \centering
    \includegraphics[width=0.37\textwidth]{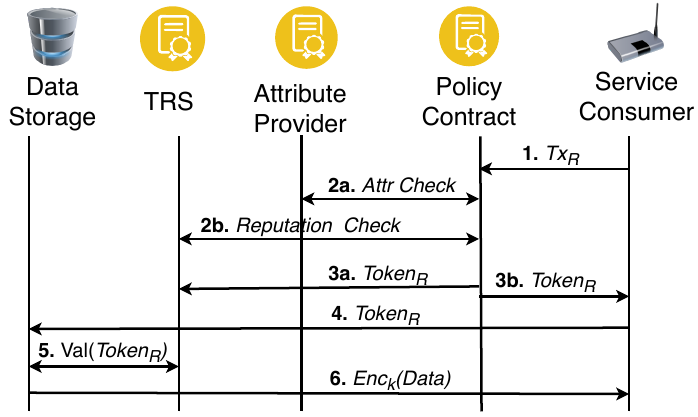}.
    \caption{Resource access via smart contract.}
    \label{fig:access-via-smart-contract}
    \end{figure}

    In contrast to access request via service provider, access via smart contract is a relaxed setting, in which $SP$ ignores Eq.~\ref{eq:access-final-decision} and relies solely on the reputation of the $SC_i$ calculated by $CTR_{TRS}$. Hence, $SC_i$ may obtain the $Token_{R}$ with lower latency. As $SP_j$ is not involved in this scheme, $SP_j$ may be offline temporarily to save energy. We illustrate the process of access request via smart contract in Fig.~\ref{fig:access-via-smart-contract}.
    
    \begin{itemize}
        \item Step 1: Unlike authorization via service provider, $SC$ directly initiates $TX_R$ to $CTR_{pol}$. In this scenario, $TX_{R}$ is similar to Eq.~\ref{eq:tx-r} but without $Sig_{ SP_j }$.
        \item Step 2a and 2b: Policy contract validates $TX_R$ against policy $P_r$ by calling attribute check to $CTR_{AP}$ (2a) and reputation check to $CTR_{TRS}$ (2b). Policy check is performed according to Algorithm~\ref{alg:access-control-mechanism}.
        \item Step 3a and 3b: Upon successful validation, $CTR_{pol}$ issues $Token_R$ by storing it in $CTR_{TRS}$ (3a) and sending it to $SC_i$. At this stage, $SC_i$ is authorized to access resource $r$. If the validation fails, $CTR_{pol}$ records it to $CTR_{TRS}$.
        \item Step 4, 5, and 6: To collect the data from resource $r$, $SC_i$ sends $Token_R$ to the dedicated data storage. The data storage validates $Token_R$ to $CTR_{TRS}$ to prevent access token forgery. Upon successful validation, data storage sends ${Enc}_{k}(Data)$, data encrypted by session key $k$ to $SC_i$.
    \end{itemize}
    
    \subsubsection{Handling Malicious Behavior}
    \label{subsub:events-generation}
    Recall in our definition of trust and reputation that $SP$ is responsible for keeping track and progressively evaluating the trust value across different nodes, while $CTR_{TRS}$ is in charge of progressively evaluating and storing reputation values across participating nodes. In the authorization process, $CTR_{pol}$ validates if $SC$ violates access policy $P$. Meanwhile, in the access token validation process, dedicated data storage checks if the $Token_R$ is invalid. Any violations are reported as a malicious behavior to $CTR_{TRS}$ and recorded securely on the blockchain. $CTR_{pol}$ then initiates an event by which $SP$ uses to update the reputation and blacklist the corresponding node. Note that, all $SP$s are notified of malicious activities by the $CTR_{pol}$.

\section{Performance Evaluation}
\label{sec:performance-evaluation}
This section describes the proof-of-concept implementation of our proposed architecture in a private Ethereum network. To illustrate the feasibility of our architecture, we divide the performance evaluation into four parts: (1) delay analysis, (2) smart contract evaluation, (3) reputation evolution, and (4) trust value evolution.

    \subsection{Implementation}
    \label{sub:implementation}
    We implemented our proposed architecture on a MacBook Pro laptop (8GB RAM, 2.3 GHz Intel Core i5 CPU, MacOS 10.14.6). We built a private Ethereum network with single miner using Docker containers\footnote{\url{https://www.docker.com/products/docker-desktop}} to simulate a real-world network, in which each peer and miner run \texttt{geth} v1.9.6 to manage the network. We used Python v3.7.4 scripts to simulate the interactions among nodes, \texttt{web3} v5.2.0 library for communicating with Ethereum peers, and \texttt{py-solc} v3.2.0 library for compiling the smart contracts. To mimic resource constraints in IoT devices, we limit the CPU cycles of all peers to $0.25$ (25\% of available processing time)~\cite{dockerlimit}.
    
    All smart contracts are written in Solidity v0.5.11~\cite{solidity0511}, a Turing complete object oriented language for Ethereum Virtual Machine (EVM). Solidity has native support for computing and verifying hashes efficiently~\cite{8752021}. However, complex function execution in Solidity may cost a considerable amount of Gas. Moreover, Solidity v0.5.11 only supports basic computation in integers and does not support floating point computation~\cite{solidity0511}. We changed Eq.~\ref{eq:reputation-formula} so that ${ { Rep }_{ i }\in \mathbb{Z} }$, i.e., all reputations are stored as integers.
    
    As blockchain is immutable, the smart contract update mechanism works by flagging the old smart contract as obsolete\footnote{In Ethereum, it is performed by the predefined \texttt{selfdestruct} method.}, and deploying a new smart contract. To facilitate secure smart contract updates, we separate logic and data into different smart contracts. The main smart contract, which contains all functions for the access control and TRS logic, does not store data and connects to the secondary contract to collect and store data. This way, when a \texttt{selfdestruct} method is called and a new smart contract is deployed, the old smart contract becomes unusable but the old data remains accessible to the new smart contract. This is important when a bug is discovered and an update is necessary.
    
    Note that we selected private Ethereum blockchain as the implementation platform to show that our solution works for both permissioned and permission-less blockchains. However, our proposed architecture can be implemented in any blockchain instantiation that supports smart contract execution, for instance Hyperledger Fabric and Sawtooth.

    \begin{table*}[t]
    \centering
    \caption{A summary of gas used and elapsed time for contract deployments and function calls.}
    \label{tab:gas-used-elapsed-time}
    \begin{tabular}{llcS[table-format=6]S[table-format=6]SSS[table-format=4.2]S[table-format=4.2(3)]}
    \toprule
    \multicolumn{2}{c}{\textbf{Methods}} & \multirow{2}{*}{\textbf{Type}} & \multicolumn{3}{c}{\textbf{Gas Used}} & \multicolumn{3}{c}{\textbf{Elapsed Time (ms)}} \\
    \multicolumn{1}{c}{Method} & \multicolumn{1}{c}{Contract} &  & \multicolumn{1}{c}{Min}    & \multicolumn{1}{c}{Max}    & \multicolumn{1}{c}{Average}    & \multicolumn{1}{c}{Min}      & \multicolumn{1}{c}{Max}     & \multicolumn{1}{c}{Average}     \\ \midrule
    - & $CTR_{AP}$ & Ctr. depl. & {-} & {-} & 655608.00(0) & 1332.3 & 18574.65 & 5724.71(408)K \\
    - & $CTR_{pol}$ & Ctr. depl. & {-} & {-} & 1654067.00(0) & 1331.91 & 12074.71 & 5078.07(283)K \\
    - & $CTR_{TRS}$ & Ctr. depl. & {-} & {-} & 1036234.00(0) & 1158.60 & 14877.57 & 5367.94(312)K \\
    \texttt{regAttr} & $CTR_{AP}$ & Func. call & 159014 & 174078 & 159575.86(269)K & 1031.16 & 9150.20 & 4784.21(189)K \\
    \texttt{regPolicy} & $CTR_{pol}$ & Func. call & 303422 & 318422 & 303922.00(269)K & 605.43 & 25747.86 & 5795.94(538)K \\
    \texttt{recMisbehavior} & $CTR_{TRS}$ & Func. call & 100055 & 195334 & 103230.97(171)K & 1580.69 & 12177.24 & 5092.43(265)K \\ \bottomrule
    \end{tabular}
    \end{table*}

    \subsection{Results}
    \label{sub:results}
    
        \subsubsection{Delay Analysis}
        \label{subsub:delay-analysis}
\begin{figure}[!t]
        \centering
        \includegraphics[width=0.47\textwidth]{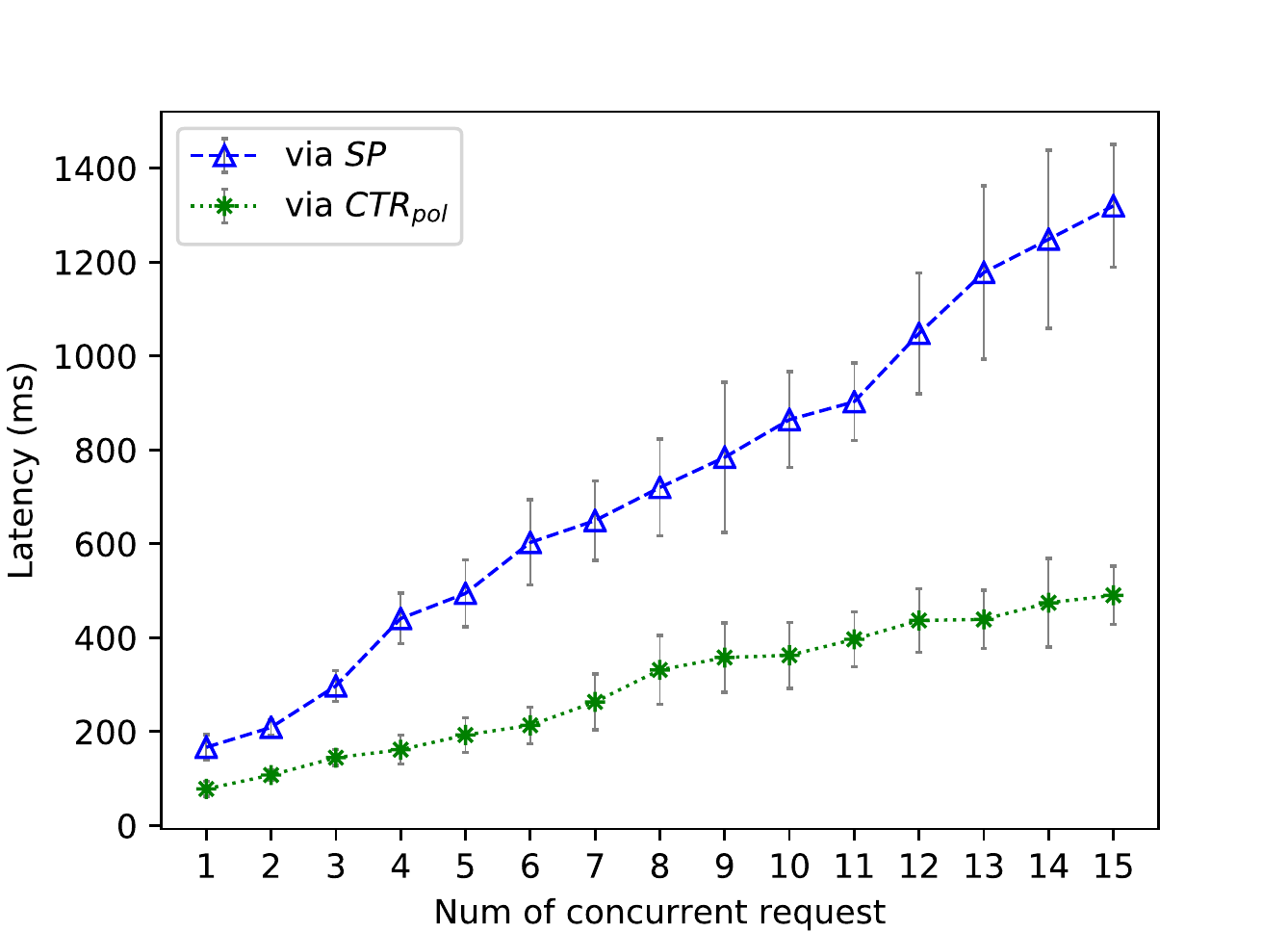}
        \caption{Elapsed time for authorization.}
        \label{fig:latency}
        \end{figure}

        Recall that in Section~\ref{sub:access-control-scheme}, an $SC$ may request an access via $SP$ or $CTR_{pol}$ smart contract. In this evaluation, we are interested in seeing the effect of having multiple access requests simultaneously for both via an $SP$ or $CTR_{pol}$. Latency is the time required for an $SC$ to be authorized. We conducted an experiment in which up to 15 concurrent requests are performed. We repeated the experiment 30 times. We plot the result in Fig.~\ref{fig:latency}.
        
        As described in Section~\ref{sub:access-control-scheme}, authorization request via $SP$ involves more steps, in which the $SP$ must acknowledge and finalize each request by executing Eq.~\ref{eq:access-final-decision} and signing $Token_{R}$ using its private key. This incurs higher delay for $SP$ authorization requests. On the other hand, authorization request via $CTR_{pol}$ reduces delay as the requests are directly routed to the blockchain. This results in longer latency for request via $SP$, especially when the number of concurrent requests are high. The rate of increase in latency for contract-based authorisation is much lower than the rate for SP-based authorisation, showing that the smart-contract approach increases scalability. 
        
        \subsubsection{Smart Contract Evaluation}
        \label{subsub:contract-deployment}
        We measured the required gas and elapsed time for deploying each mandatory smart contract and calling three primary functions in each smart contract: (1) \texttt{regAttr} for attribute registration, (2) \texttt{regPolicy} for policy registration, and (3) \texttt{recMisbehavior} for misbehavior logging. In Ethereum, Gas cost depends on the number of EVM opcodes involved when executing a function~\cite{wood2014ethereum}. We repeated the experiment 30 times and we summarize the result in Table~\ref{tab:gas-used-elapsed-time}.
        
        The gas costs for smart contract deployments remain the same for all repeated measurements whilst the gas costs for each function call varied. Note that, gas costs are determined by the number of low level operation (opcodes) required to execute a function in the Ethereum Virtual Machine (EVM). Each opcode is associated with a gas cost, e.g., 3 gas for \texttt{ADD} (addition) and 5 gas for \texttt{MUL} (multiplication). As the arguments in each function call differ in size, which results in the variation of EVM opcodes, the gas costs are dynamic. However, as each contract deployment does not incur variation in EVM opcodes, so the gas costs remain static. Note that, all costs associated with executing all functions are the responsibility of $SC$s.
        
        The elapsed time varies for both contract deployments and function calls. The fastest deployment time is approximately 600 milliseconds, while the worst case may take up to 25 seconds.
        The elapsed time is measured from initiating the transaction until the transaction is stored permanently on the blockchain. In Ethereum, the signed transaction first goes to the transaction pool prior to being appended to a block by a miner. Note that there is no precise timing guarantee associated with when a miner picks the signed transaction and appends it to the blockchain. The uncertainty in this process causes variations in elapsed time, as the signed transaction may be directly picked and stored on the blockchain or stay in the transaction pool until the next block generation.
        
        \begin{figure}[!t]
        \centering
        \includegraphics[width=0.47\textwidth]{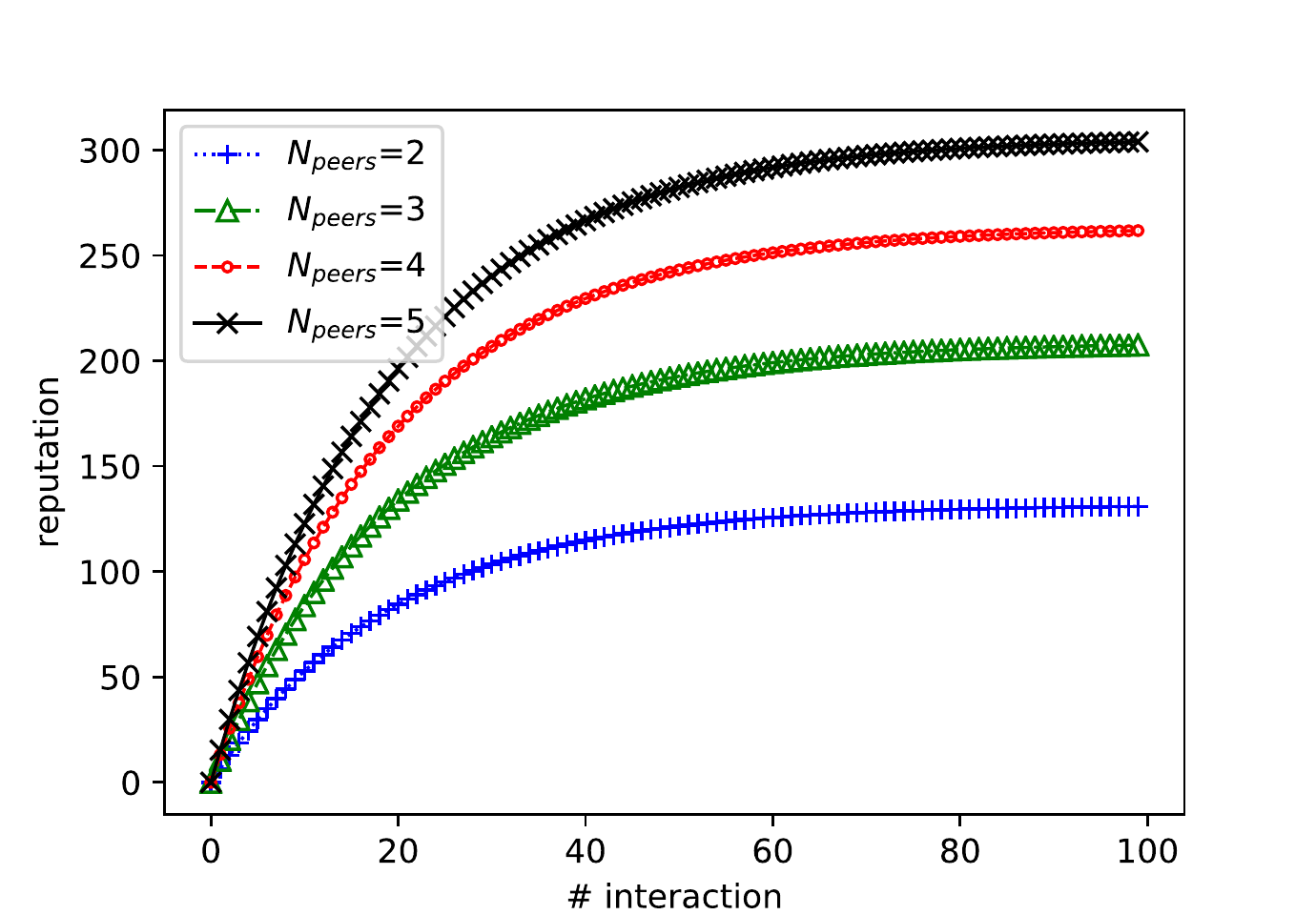}
        \caption{Reputation evolution.}
        \label{fig:reputation-evolution}
        \end{figure}
        
        \subsubsection{Reputation Evolution}
        \label{subsub:reputation-evolution}
        To observe the trends in reputation growth, we simulated benign node behavior and calculated its trust value using Eq.~\ref{eq:reputation-formula} with different values of $N_{peers}$. We set $\beta_{pos}$ and $\lambda$ to $10$ and $0.95$, respectively. Recall that, in Eq.~\ref{eq:reputation-formula} $\beta$ sets the increase and decrease for each positive and negative interaction and $\lambda$ sets the growth rate or the $y$ scaling. We particularly set $\lambda$ closer to $1$ to achieve gradual growth. We plot the result in Fig.~\ref{fig:reputation-evolution}.
        
        As shown in Fig.~\ref{fig:reputation-evolution} all reputation increases follow logarithmic growth and saturate proportionally at a certain value for all values of $N_{peers}$. The saturation point depends on $N_{peers}$, i.e., higher $N_{peers}$ causes saturation at higher reputation value. In this scenario, reputation is designed to have no upper limits, as widely used in online service provision~\cite{Jøsang2007}. On the other hand, trust value ranges from $0$ to $1$.
    
        \subsubsection{Trust Evolution}
        \label{subsub:trust-evolution}
\begin{figure}[!t]
        \centering
        \includegraphics[width=0.47\textwidth]{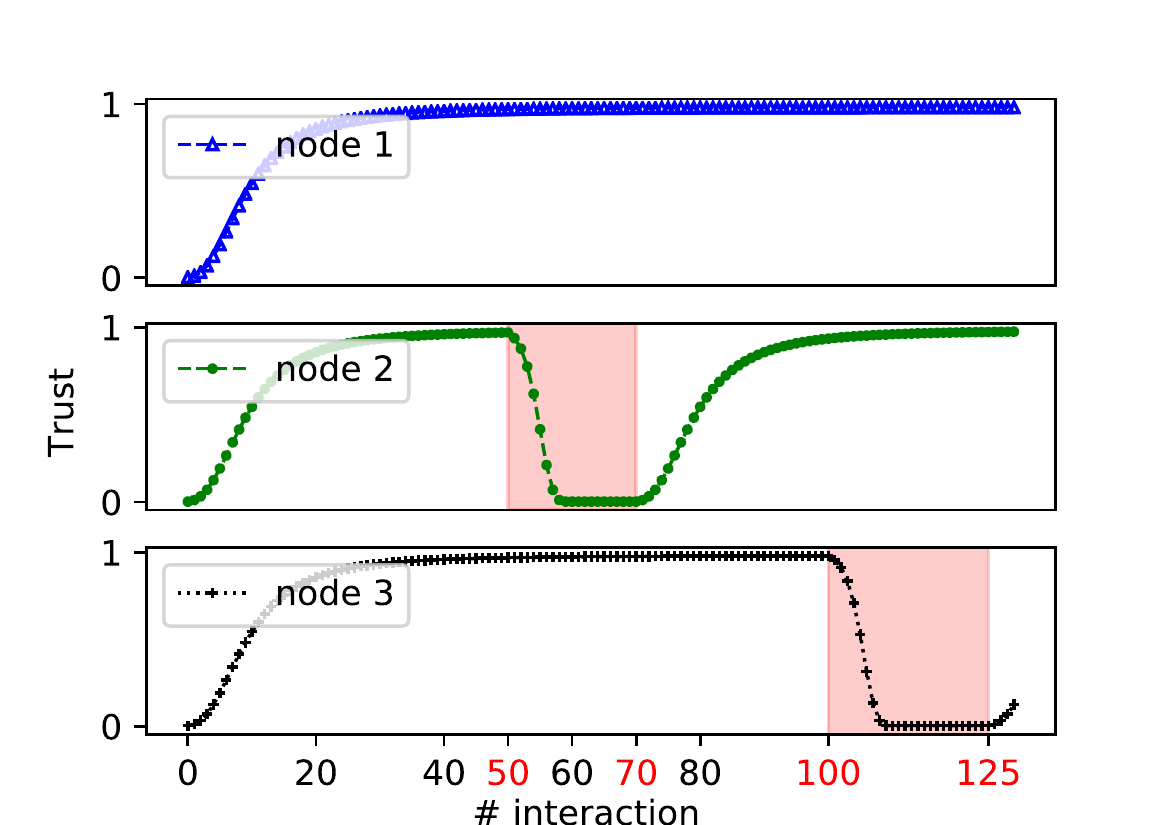}
        \caption{Trust value evolution of 3 nodes trying to access $r_A$ of $SP_A$.}
        \label{fig:trust-evolution}
        \end{figure}

        To investigate the evolution of trust value, we simulated three nodes accessing a resource $r_A$ of $SP_A$, two of which act maliciously, i.e., using incorrect access token, during a certain time epoch (\# of interaction) over the total experiment length of 130 interactions. Node 2 begins to act maliciously from interaction 50 to 70, while node 3 from interaction 100 to 125. $SP_{A}$ progressively calculated the trust value using both Eq.~\ref{eq:ageing-function} and Eq.~\ref{eq:gompertz-function} in each interaction. To set a gradual incline and to limit trust from $0$ to $1$, we set the value of $\gamma$, $a$, $b$, and $c$ to $0.95$, $1$, $-6$, and $-0.1$, respectively. We plot the results in Fig.~\ref{fig:trust-evolution}.
        
        To reach significant trust value, a node needs at least 30 to 40 benign interactions. On the other hand, losing significant trust value only requires less than 10 malicious interactions, i.e., see the shaded area in Fig.~\ref{fig:trust-evolution}.
        Note that, as the trust growth is modeled using Gompertz function, the trust will never reach its maximum value, i.e., $T_{max}=1$.
        
        As calculated by $CTR_{TRS}$, the global reputation score of node 2 and node 3 drop significantly, as an implication of their malicious actions. Subsequently, $CTR_{pol}$ emits malicious behavior events to notify other $SP$s about the malicious actions of node 2 and node 3.
        
\section{Discussion}
\label{sec:discussion}
In our architecture, the adversaries are $SC$s who can perform attacks related to access control, i.e., successively requesting access and manipulating access tokens to collect data using forged and expired tokens or to leverage access rights. We assume that the $SC$s are not capable of performing physical attacks that would damage any device. We further assume that $SP$s are mainly controlled by an honest and reliable building manager.
The $SP$s are motivated to protect their resources from unwanted access. So they perform signature and access token validation accordingly. In future work, we plan to investigate the case where $SP$s can be malicious or unreliable in data delivery.
All miners are bounded by cryptographic primitives of the blockchain specifications, which prevents malicious miners to take control over the blockchain network. 
We adopt the access control matrix~\cite{sandhu1994access} to model access control. The model defines a set of subjects, a set of objects, a set of resources, and a set of actions, which correspond to $SC$, $SP$, $r$, and $\tau$ in our scheme, respectively.
We further assume that an anomaly detection system is already present to protect the network.

In case where a node $A$ wants to ruin node $B$'s reputation, i.e., launch a bad mouthing attack, the request signature proves the originality of the sender. Hence a node cannot simply use tokens of other nodes to maliciously drop its reputation. Our proposed scheme is also resilient to sybil attack, i.e., a node trying to forge new identities to its advantage, and newcomer attacks, i.e., a node trying to get a new identity to reset its bad reputation. Both sybil and newcomer attacks are prevented by the attribute registration scheme, where a node cannot re-register itself to the attribute authority, as the latter records the attributes relation with the actual device information.

In our framework, trust and reputation are computed using different formulas, i.e., Eq.~\ref{eq:ageing-function} and~\ref{eq:gompertz-function} for trust, and Eq.~\ref{eq:reputation-formula} for reputation, and also calculated by different instance, we expect that trust and reputation values for a single $SC$ to vary across the network. An implication of this is the possibility that $SP_x$ does not incorporate negative interaction of $SC_z$ to $SP_y$ in its trust calculation. However, this can be mitigated by defense mechanisms explained in Section~\ref{subsub:events-generation}.

Our proposed framework has a few limitations. Recall in Section~\ref{sub:implementation}, we separate logic and data implementation for the smart contract. This scheme may help to enforce secure smart contract upgrade, as the logic for smart contract can be updated by marking a contract obsolete and deploying a new contract, if bugs are discovered. However, this scheme does not allow data structure updates. If the contract is updated, the logic may change but the data structure remains the same. Another limitation is that our framework only supports IoT devices which have sufficient computational power to perform asymmetric cryptography.



\section{Conclusion}
\label{sec:conslusion-future-work}
In this paper, we proposed a trust management framework for decentralized IoT access control system, which aims to provide secure and trustworthy access control while being able to detect and eliminate malicious and compromised nodes.
We incorporated trust and reputation in our proposed access control mechanism, which are progressively evaluated in each interaction.
Our conceptual framework is blockchain agnostic and can be implemented in any blockchain instantiation that has sufficient support for smart contract execution.
We implemented a proof-of-concept implementation in a private Ethereum network of docker containers to benchmark our proposed architecture. Experimental results show that our proposed framework only incurs small processing delays and is feasible for trust management in decentralized IoT access control.


\bibliographystyle{./bibliography/IEEEtran}
\bibliography{./bibliography/mybib}

\end{document}